\documentclass{iopart}
\usepackage{iopams}
\usepackage{epsfig}

\newcommand {\zz}{ZrZn$_2$ }

\newcommand {\I}{\rmi}

\renewcommand {\vec}{\bi}

\begin{document}

\letter{Competition between disorder and exchange splitting in superconducting \zz}
\author{B J Powell\footnote[1]{Present address: Department of Physics,
University of Queensland, Brisbane, Qld 4072, Australia.}, James F
Annett and B L Gy\"orffy}
\address{H H Wills Physics Laboratory, University of Bristol, Tyndall Avenue,
Bristol BS8 1TL, UK}

\ead{powell@physics.uq.edu.au}

\begin{abstract}
We propose a simple picture for the occurrence of
superconductivity and the pressure dependence of the
superconducting critical temperature, $T_{SC}$, in ZrZn$_2$.
According to our hypothesis the pairing potential is independent
of pressure, but the exchange splitting, $E_{xc}$, leads to a
pressure dependence in the (spin dependent) density of states
(DOS) at the Fermi level, $D_\sigma(\varepsilon_F)$. Assuming
p-wave pairing $T_{SC}$ is dependent on $D_\sigma(\varepsilon_F)$
which ensures that, in the absence of non-magnetic impurities,
$T_{SC}$ decreases as pressure is applied until it reaches a
minimum in the paramagnetic state. Disorder reduces this minimum
to zero, this gives the illusion that the superconductivity
disappears at the same pressure as ferromagnetism does.
\end{abstract}

\submitto{\JPCM}



The coexistence of ferromagnetism and superconductivity is a
problem of long standing and general interest. Thus its recent
discovery in UGe$_2$ \cite{Saxena}, \zz \cite{Pfleiderer} and
URhGe \cite{URhGe} is attracting considerable attention. In
particular its occurrence in \zz is intriguing because, at ambient
pressure, \zz is a weak ferromagnet ($T_{FM}\approx 28.5$~K) and
by the application of pressure it can be tuned through a quantum
critical point (QCP) ($P_C\approx 21$~kbar) to become a
paramagnetic metal \cite{Pfleiderer}. This revives an old
suggestion of Fay and Appel \cite{Fay&Appel}. These authors
calculated $T_{SC}$ mediated by paramagnons in a McMillan like
formalism and found that there is superconductivity in the
(triplet) A$_1$ channel \cite{Vollhardt} on both sides of the QCP.
However, while the broad description of Fay and Appel agrees with
the observations the fine details do not. In Fay and Appel's
theory the transition temperature goes to zero at the QCP,
$T_{SC}$ then rises to a (local) maxima as the model is tuned away
from criticality (experimentally this corresponds to pressure
being varied away from $P_C$). $T_{SC}$ then falls again away from
the QCP. They also predicted that $T_{SC}$ would be approximately
the same magnitude in both the ferromagnetic and paramagnetic
sides of the phase diagram (although slightly higher on the
paramagnetic side).

When superconductivity was observed in \zz it was only seen on the
ferromagnetic side of the phase diagram and further the maximum in
$T_{SC}$ was observed at ambient pressure \cite{Pfleiderer}. The
experiments show a monotonic decrease in $T_{SC}$ with pressure
until about 18 kbar. No data has been published in the pressure
range 18 kbar $< P <$~22 kbar, therefore one cannot ascertain
where in this range $T_{SC}$ falls to zero. Several groups have
attempted to explain this either by revisiting the theory of Fay
and Appel and examining specific coupling mechanisms
\cite{Kirkpatrick_etal} or by considering the problem within a
Ginzburg--Landau formalism \cite{Walker&Samokhin}. Both of these
groups predicted that $T_{SC}$ goes to zero at $P_C$. Here we will
present a simple alternative to these scenarios for the
coexistence of superconductivity and ferromagnetism in ZrZn$_2$.
We will show that no variation in the coupling constant with
pressure (i.e. proximity to the QCP) is required to explain the
experiments due to the natural enhancement of the A$_1$ phase
transition temperature in the ferromagnetic phase and the extreme
sensitivity of the triplet pairing states to scattering from
non-magnetic impurities. On the other hand our arguments rely on
\zz being a rare example of a Stoner ferromagnet. Interestingly in
our consideration the QCP plays no special role and the pressure
at which superconductivity disappears is predicted to be strongly
sample dependent.

We wish to consider the problem via the simplest model which has
the possibility of illustrating the relevant physical phenomena:
triplet superconductivity and ferromagnetism. To this end we study
a one band Hubbard model with an effective, attractive, pairwise,
nearest neighbour interaction, $U_{ij\sigma\sigma'}$. We allow
ferromagnetism to enter via the Stoner model which appears to be
in good agreement with the observed behaviour of the ferromagnetic
phase of \zz \cite{Pfleiderer,Cordes,Uhlarz,Yates}. Thus we study
the consequences of the following Hamiltonian:

\begin{equation} 
\hat{\cal{H}}=
-\sum_{ij\sigma}t_{ij}\hat{c}_{i\sigma}^\dag\hat{c}_{j\sigma}
+\frac{1}{2}\sum_{ij\sigma\sigma'}U_{ij\sigma\sigma'}\hat{c}_{i\sigma}^\dag\hat{c}_{i\sigma}\hat{c}_{j\sigma'}^\dag\hat{c}_{j\sigma'}
- \sum_{i\sigma}\sigma E_{xc} \hat{c}_{i\sigma}^\dag
\hat{c}_{i\sigma}
\end{equation}
where $\hat{c}_{i\sigma}^{(\dag)}$ are the usual annihilation
(creation) operators for electrons with spin $\sigma = \pm 1$
occupying a tight binding orbital centred on the lattice site
labelled by $i$. To render the model tractable we assume that the
sites $i$ form a simple cubic lattice.

For a general triplet pairing state in a field gap equations
cannot be derived in the same way as they can for the zero field
case \cite{Sigrist&Ueda,BenThesis}. However, if we specialise to
the case of equal spin pairing (ESP) and neglect the action of the
dipolar field on the orbital motion of the electrons a remarkable
simplification occurs as shown in reference \cite{BenThesis} the
gap equations are

\begin{eqnarray}
\Delta_{\sigma\sigma}({\vec{k}}) = - \sum_{\vec{k}'}
\frac{U_{\sigma\sigma}(\vec{k}-\vec{k}')
\Delta_{\sigma\sigma}({\vec{k}'})}{2 E_\sigma(\vec{k}')}
(1-2f_{E_{\vec{k}'\sigma}}) \label{eqn:gap eqn}
\end{eqnarray}
where the quasiparticle spectrum is given by

\begin{eqnarray}
E_{\vec{k}\sigma} = \sqrt{\left( \varepsilon_{\vec{k}} - \mu -
\sigma E_{xc} \right) ^2 + |\Delta_{\sigma\sigma}({\vec{k}})|^2}.
\label{eqn:simple spectrum}
\end{eqnarray}

Equations (\ref{eqn:gap eqn}) and (\ref{eqn:simple spectrum}) have
several surprising features. Firstly, there is a complete
decoupling of the two spin states; even in the presence of Cooper
pairing. Secondly, the exchange splitting enters only in the role
of a chemical potential, but with opposite signs for the two spin
states. It must be stressed that these results are only valid for
ESP states. However, the large exchange splitting in a ferromagnet
probably precludes opposite spin pairing (OSP) states (that is
singlet states, $S_z=0$ triplet states or even the
Fulde--Ferrell--Larkin--Ovchinnikov (FFLO) state, which is only
stable for moderate exchange splittings). In short \eref{eqn:gap
eqn} and \eref{eqn:simple spectrum} are just the
Hartree--Fock--Gorkov approximations restricted to the ESP sector
of the theory. For
$U_{ij\sigma\sigma'}=U_{ij}\delta_{\sigma\sigma'}$ no such
restriction is required as OSP states are not a possibility.

As $T \rightarrow T_{SC}$, $|\Delta_{\sigma\sigma'}(\vec{k})|
\rightarrow 0$ and we find the linearised gap equations:

\begin{figure}
    \centering
    \epsfig{figure=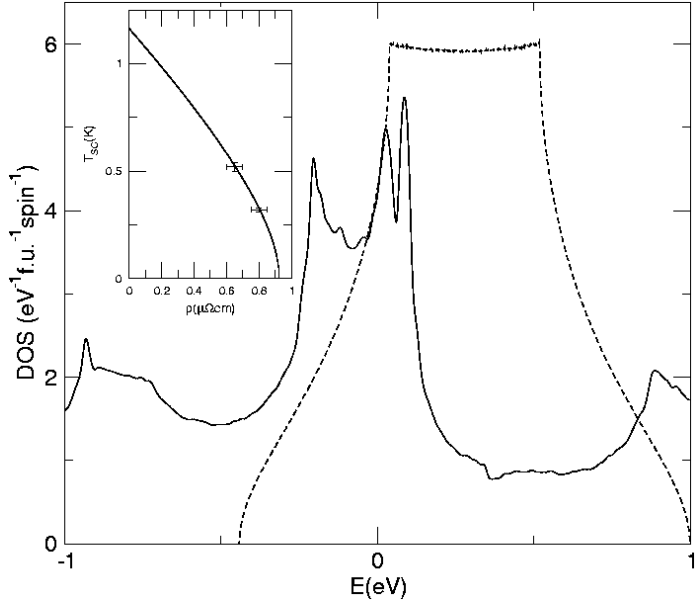, height=8cm, angle=0}
    \caption{The DOS
    from the LDA calculations for \zz (\full) by
    Santi and coworkers \cite{Yates,Santi} and our tight binding fit to the LDA DOS (\broken).
    Inset: The experimental data \cite{private_steves} is consistent
    with a fit to
    $T_{SC}^0(P=0)\sim 1.2$~K from the Abrikosov--Gorkov formula (\ref{eqn:AG}).} \label{fig:zrzn2_DOS}
\end{figure}

\begin{eqnarray} \fl
\Delta_{\sigma\sigma}({\vec{k}}) = \sum_{\vec{k}'}
\frac{U_{\sigma\sigma}(\vec{k}-\vec{k}') }{2 \big( \varepsilon
(\vec{k}') -\mu - \sigma E_{xc} \big)} \tanh \left(
\frac{\varepsilon (\vec{k}') -\mu - \sigma E_{xc}}{2k_BT} \right)
\Delta_{\sigma\sigma}({\vec{k}'}). \label{eqn:linearised gap eqn}
\end{eqnarray}
Before solving these equations numerically we must first choose
the hoping integrals, $t_{ij}$, and the coupling constants
$U_{ij\sigma\sigma'}$. We fit the hoping integrals with an on site
($t_{ii}=\mu$) and nearest neighbour terms only so as to give the
same relative density of states in the region of the Fermi level
as is found in \textit{ab initio} band structure calculations
\cite{Yates,Santi}. The DOS from our fit is compared with that
found in the \textit{ab initio} calculation in figure
\ref{fig:zrzn2_DOS}. Evidently, our one band model cannot
reproduce the complex behaviour of the DOS over the full energy
range of the Zr related d-band, nevertheless, within an energy
range of 20~meV about the Fermi energy it does do so. Since
$U_{ij\sigma\sigma'}$ depends on only one parameter: $U_{ij} = U$
for sites $i$ and $j$ being nearest neighbours ($U_{ij}=0$
otherwise) it can be determined by reference to the measured
$T_{SC}$ for clean ZrZn$_2$. This is hampered by the extreme
sensitivity of triplet pairing to scattering from non-magnetic
impurities \cite{Larkin,Mackenzie_Tc} and by the lack of data.
Using experimental estimates of the residual resistivity, $\rho$,
we find that what little data there is \cite{private_steves} is
consistent with a clean superconducting critical temperature, at
ambient pressure, $T_{SC}^0(P=0) \sim 1.2$~K as shown in the inset
to \fref{fig:zrzn2_DOS}.

We solved (\ref{eqn:linearised gap eqn}) numerically with $U =
0.88t$ on a k-space integration mesh of $10^9$ points. Such a fine
integration mesh is required to accurately reproduce the DOS. Our
method (implicitly) requires an accurate calculation of the
$D_\sigma(\varepsilon_F)$ as we are varying the exchange splitting
and thus we are changing $D_\sigma(\varepsilon_F)$, so any errors
in evaluating $D_\sigma(\varepsilon_F)$ will lead to significant
errors in our calculation of the variation of $T_{SC}$ with
$E_{xc}$.

\begin{figure}
    \centering
    \epsfig{figure=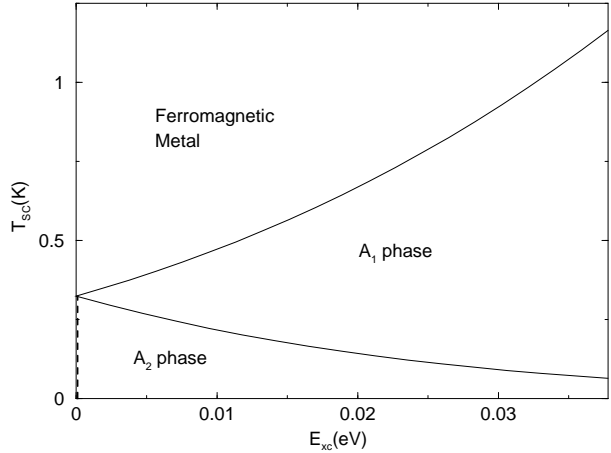, height=8cm, angle=270}
    \caption{The phase diagram of our model. The critical temperature is shown for both
    A$_1$ and A$_2$ phases over a range of exchange splittings.
    The hatched area indicates the A phase, which is the ground state when $E_{xc}=0$.}
    \label{fig:numerical_res_zz}
\end{figure}

The results of our numerical calculations are shown in
\fref{fig:numerical_res_zz}. The A$_1$ phase displays
superconductivity in only the $\uparrow\uparrow$ channel. In the
A$_1$ phase $\vec{d}(\vec{k})\sim\left( k_x + \I k_y,\I (k_x + \I
k_y),0\right)$, where $\vec{d}(\vec{k})$ is the usual BW vector
order parameter for triplet superconductivity \cite{Vollhardt}.
The A$_2$ phase corresponds to superconductivity in both ESP
channels but with different amplitudes in the two channels so that
$\vec{d}(\vec{k})\sim\left( k_x + \I k_y,\I\kappa (k_x + \I
k_y),0\right)$ where $0<\kappa<1$. The A phase, which is only
stable in zero exchange spitting, has the same pairing amplitude
for both of the ESP states, corresponding to the order parameter
$\vec{d}(\vec{k})\sim\left( k_x + \I k_y,0,0\right)$. Because of
the complete separation of the up and down sheets the linearised
gap equations \eref{eqn:linearised gap eqn} can be used to
calculate the lower transition temperature, $T_{SC}^{A_2}$. This
transition represents the formation of a superconducting state in
the minority spin state. Our phase diagram of course assumes that
no other phase transitions occur. The existence of the A$_1$ and
A$_2$ phases only is consistent with a general symmetry analysis
\cite{Mineev} and the Ginzburg--Landau expansion of the free
energy \cite{BenThesis,Ben1}.

\begin{figure}
    \centering
    \epsfig{figure=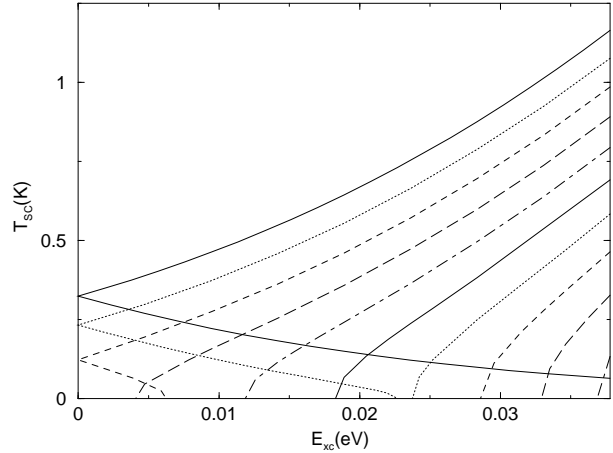, height=8cm, angle=270}
    \caption[The critical temperature of a ESP p-wave
    superconductor and the temperature of the A$_1$-A$_2$
    transition as a function of exchange splitting in the presence
    of disorder.]{The critical temperature of a ESP p-wave
    superconductor and the temperature of the A$_1$-A$_2$
    transition as a function of exchange splitting in the presence
    of disorder. The curves correspond (from the top down) to
    $\rho_{tr}$ = 0, 0.1~$\mu\Omega$cm, 0.2~$\mu\Omega$cm,
    0.3~$\mu\Omega$cm, 0.4~$\mu\Omega$cm, 0.5~$\mu\Omega$cm,
    0.6~$\mu\Omega$cm, 0.7~$\mu\Omega$cm, 0.8~$\mu\Omega$cm and
    0.9~$\mu\Omega$cm.} \label{fig:tc_e_ref}
\end{figure}

To make contact with experiment we must allow for the strong
dependence of the superconducting transition temperature of a
p-wave superconductor on non-magnetic impurity scattering
\cite{Larkin,Mackenzie_Tc}. This is done via the Abrikosov--Gorkov
formula:

\begin{equation}
\ln \left(\frac{T_{SC0}}{T_{SC}} \right) = \psi\left( \frac{1}{2}
+ \frac{\hbar}{4\pi\tau_{tr} k_BT_{SC}} \right) - \psi\left(
\frac{1}{2} \right) \label{eqn:AG}
\end{equation}
where $\tau_{tr}$ is the quasiparticle lifetime as measured in
transport experiments and $\psi(x)$ is the digamma function. Note
that we do not need to worry about the Baltensperger--Sarma
equation \cite{Baltensperger,Sarma}, which accounts for the
reduction in the critical temperature of a superconductor due to
exchange splitting, as this is only valid for OSP states. Thus,
the Abrikosov--Gorkov formula can be used to calculate $T_{SC}$ as
a function of $\tau_{tr}$, or equivalently $\rho$ via the Drude
formula. To make the most of the available experimental data we
used the Abrikosov--Gorkov formula \eref{eqn:AG} to investigate
the effect of disorder on the above phase diagram (see
\fref{fig:tc_e_ref}).

The final step needed to make a direct comparison with
measurements of $T_{SC}$ as a function of pressure is to note
that, experimentally, the Curie temperature is a linear function
of pressure \cite{Cordes,Uhlarz}. Namely, $T_{FM}(P) = T_{FM}(0)
\left( 1 - P/P_C \right)$. The zero temperature magnetisation is
also linear in pressure \cite{Cordes} and thus proportional to
$T_{FM}$ (as is predicted by the Stoner model), giving $M(P,T=0) =
M(0,0) \left( 1 - P/P_C \right)$. For a Stoner ferromagnet, the
magnetisation is linearly dependent on the exchange splitting and
hence

\begin{eqnarray}
E_{xc}(P,T=0) = \left\{
\begin{array}{ll}
E_{xc}(0,0) \left( 1 - \frac{P}{P_C} \right) & P \leq P_C \\
0 & P > P_C.
\end{array} \right.
\end{eqnarray}

\begin{figure}
    \centering
    \epsfig{figure=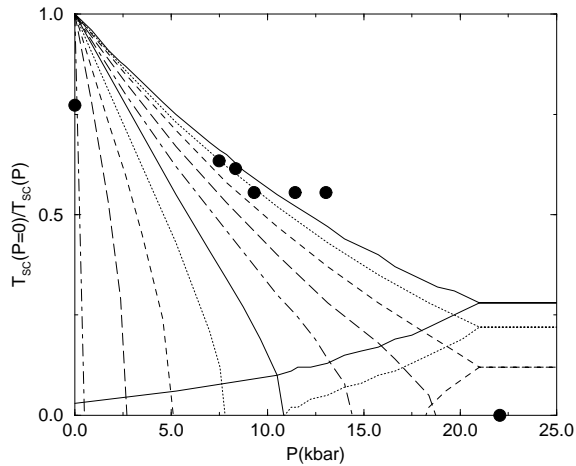, width=8.5cm, angle=0}
    \caption{The critical temperature of a ESP p-wave
    superconductor and the temperature of the A$_1$-A$_2$
    transition as a function of pressure in the presence
    of disorder. The theoretical curves are scaled so that $T_{SC}=1$
    at ambient pressure. The experimental data, taken from
    \cite{Pfleiderer},
    was scaled in the same way after straight line had been fitted to the data.
    The curves correspond (moving from top right to bottom left) to
    $\rho_{tr}$ = 0, 0.1~$\mu\Omega$cm, 0.2~$\mu\Omega$cm,
    0.3~$\mu\Omega$cm, 0.4~$\mu\Omega$cm, 0.5~$\mu\Omega$cm,
    0.6~$\mu\Omega$cm, 0.7~$\mu\Omega$cm, 0.8~$\mu\Omega$cm and
    0.9~$\mu\Omega$cm.} \label{fig:tc_P}
\end{figure}

We now invoke the fact that $T_{FM} \gg T_{SC}$ which implies that
$E_{xc}(P,T=T_{SC}) \sim E_{xc}(P,T=0)$. Thus we can map the
results of $T_{SC}(E_{xc})$ (shown in figure \ref{fig:tc_e_ref})
onto $T_{SC}(P)$ which we show in figure \ref{fig:tc_P}. It can be
seen that although quantitative agreement with experiment is not
achieved, the general features of experiment are reproduced by
several of the curves.

Thus we conclude that we have demonstrated the viability of the
following simple picture. Irrespective of the mechanism of pairing
and exchange splitting \zz is a p-wave superconductor with a low
$T_{SC} \sim 1.2$~K (at ambient pressure). This superconductivity
is not observed in the paramagnetic phase of currently available
samples due to disorder. However, the exchange field enhances
$T_{SC}$ of an A$_1$ p-wave state and this is the cause of the
observed superconductivity in the ferromagnetic phase. Experiment
suggests \cite{Pfleiderer,Cordes,Uhlarz,Yates} \zz is a rare
Stoner ferromagnet for which the exchange splitting is
proportional to the magnetic order parameter. Thus when $P>P_C$
and therefore $T_{FM}=M=E_{xc}=0$ there is no measurable
superconductivity. However, improvement in sample quality will
lead to a lowering of the residual resistivity and thus presents
the possibility of the observation of superconductivity in the
paramagnetic state (as is demonstrated by the curves with
$\rho_{tr} < 0.3~\mu\Omega$cm in \fref{fig:tc_P}). This
explanation is consistent with and lends microscopic support to
the more phenomenological arguments of Walker and Samokhin
\cite{Walker&Samokhin} and Mineev \cite{Mineev}.

\vspace*{10pt} It is a pleasure to thank Stephen Hayden, Stephen
Yates and Gilles Santi for sharing their results with us and for
helpful discussions. We would also like to thank the Laboratory
for Advanced Computation in the Mathematical Sciences
(http://lacms.maths.bris.ac.uk) for extensive use of their beowulf
facilities. One of the authors (BJP) was supported by an EPSRC
studentship and, in the final stages of this work, by the
Australian Research Council.

\section*{References}


\begin{thebibliography}{99}
\bibitem{Saxena} Saxena S S, Agarwal P, Ahilan K,
Grosche F M, Haselwimmer R K W, Steiner M J, Pugh E, Walker I R,
Julian S R, Monthoux P, Lonzarich G G, Huxley A, Sheikin I,
Braithwaite D and Flouquet J 2000 \emph{Nature} {\bf 406} 587

\bibitem{Pfleiderer}
Pfleiderer C, Uhlarz M, Hayden S M, Vollmer R, von~L{\"o}hneysen
H, Bernhoeft N R and Lonzarich G G 2001 \emph{Nature} {\bf 412} 58

\bibitem{URhGe}
Aoki D, Huxley A, Ressouche E, Braithwaite D, Flouquet J, Brison
J-P, Lhotel E and Paulsen C 2001 \emph{Nature} {\bf 413} 613

\bibitem{Fay&Appel}
Fay D and Appel J 1980 \PR B {\bf 22} 3173

\bibitem{Vollhardt}
Vollhardt D and W{\"o}lfe P 1990 \emph{The Superfluid Phases of
Helium 3} (London: Taylor and Francis)


\bibitem{Kirkpatrick_etal}
Kirpartrick T R, Belitz D, Vojta T and Narayanan R 2001 \PRL {\bf
87} 127003

\bibitem{Walker&Samokhin}
Walker M B and Samokhin K V 2002 \PRL {\bf 88} 207001

\bibitem{Cordes}
Cordes H G, Fischer K, and Pobell F 1981 \emph{Physica} B {\bf
107} 531

\bibitem{Uhlarz}
Uhlarz M, Pfleiderer C, von~L{\"o}hneysen H, Hayden S M and
Lonzarich G G 2002 \emph{Physica} B {\bf 312-313} 487

\bibitem{Yates}
Yates S J C, Santi G, Hayden S M, Meeson P J and Dugdale S B
\emph{Preprint} cond-mat/0207285

\bibitem{Sigrist&Ueda}
Sigrist M and Ueda K 1991 \RMP {\bf 63} 239

\bibitem{BenThesis}
Powell B J 2002 \emph{On the Interplay of Superconductivity and
Magnetism} Ph.D. thesis, Unversity of Bristol, UK

\bibitem{Santi}
Santi G, Dugdale S B and Jarlborg T 2001 \PRL {\bf 87} 247004

\bibitem{Larkin}
Larkin A I 1965 \emph{JETP Lett.} {\bf 2} 130

\bibitem{Mackenzie_Tc}
Mackenzie A P, Haselwimmer R K W, Tyler A W, Lonzarich G G, Mori
Y, NishZaki S and Maeno Y 1998 \PRL {\bf 80} 161

\bibitem{private_steves}
Hayden S M and Yates S J C \emph{Private communication}

\bibitem{Mineev}
Mineev V P 2002 \PR B {\bf 66} 134504

\bibitem{Ben1}
Powell B J, Annett J F and Gy{\"o}rffy B L 2002 \emph{Ruthenate
and Rutheno-Cuprate Materials: Unconventional Superconductivity,
Magnetism and Quantum Phase Transitions (Springer Lecture Notes in
Physics vol~603)} ed Noce C, Cuoco M, Romano A and Vecchione A
(Heidelberg: Springer)

\bibitem{Baltensperger}
Baltensperger W 1958 \emph{Physica} {\bf 24} S153

\bibitem{Sarma}
Sarma G 1963 \emph{J. Phys. Chem. Solids} {\bf 24} 1029

\end{thebibliography}

\end{document}